\documentclass[epjc3]{svjour3}  	
\smartqed  
\journalname{Eur. Phys. J. C}
\usepackage{graphicx}
\usepackage{graphicx}				
\usepackage{amssymb}
\usepackage{amsmath}
 \pdfoutput=1
\usepackage[usenames,dvipsnames]{color}  
\usepackage{setspace}
\usepackage{braket}
\usepackage{amsmath}
\usepackage{amssymb}
\usepackage[normalem]{ulem}
\usepackage{xcolor}
\usepackage{amssymb,wasysym}
\usepackage[utf8]{inputenc} 
\usepackage[T1]{fontenc} 
\usepackage[english]{babel}
\usepackage{subfig}
\usepackage{comment,pifont}
\usepackage{physics,hyperref}
\usepackage{booktabs}
\usepackage{graphicx}
\newcommand{\no}{\scalebox{0.70}{$\times$}}
\newcommand{\yes}{$\checkmark$} 
\usepackage{xcolor}

\journalname{Eur. Phys. J. C}

\usepackage{cite}
\begin{document}
\title{A High-Precision Statistical Analysis of the SN1987A
Neutrino Temporal Distribution
}


\author{Riccardo Maria Bozza\thanksref{e1,addr1}
        \and
         Vigilante di Risi\thanksref{e2,addr2}
         \and
         Veronica Oliviero\thanksref{e3,addr1}
         \and
         Giulia Ricciardi\thanksref{e4,addr1}
         \and
         Francesco Vissani\thanksref{e5,addr3}
}

\thankstext{e1}{e-mail: riccardomaria.bozza@unina.it}
\thankstext{e2}{e-mail: dirisi.vigilante1998@ucas.ac.cn}
\thankstext{e3}{e-mail: veronica.oliviero2@unina.it}
\thankstext{e4}{e-mail: giulia.ricciardi2@unina.it}
\thankstext{e5}{e-mail: francesco.vissani@lngs.infn.it}


\institute{ Dipartimento di Fisica E. Pancini, Universit\`a di Napoli Federico II and INFN, Sezione di Napoli, Complesso Universitario di Monte Sant'Angelo, Via Cinthia, Napoli (NA), Italy \label{addr1}
           \and
           International Centre for Theoretical Physics Asia-Pacific (ICTP-AP), University of Chinese Academy of Sciences (UCAS), 100190 Beijing, China \label{addr2}
           \and
           INFN, Laboratori Nazionali del Gran Sasso, 67100 Assergi, L’Aquila (AQ), Italy\label{addr3}
}

\date{Received: date / Accepted: date}
\maketitle
\begin{abstract}
 The interpretation of the SN1987A neutrino data remains limited by significant absolute timing uncertainties across independent detectors and a long-standing tension in the angular distributions. This work presents a high-precision statistical framework that resolves these issues by formalizing the mathematical distinction between intrinsic and relative timing offsets (ITOs and RTOs). By performing a simultaneous $\chi^2$-alignment of the Kamiokande-II and Baksan data with the IMB clock, 
we systematically incorporate non-Gaussian statistical correlations, reducing the temporal uncertainty by two orders of magnitude to the sub-second level. 
The $\chi^2$-analysis shows that Baksan's absolute timestamps require an advancement of 30.4~s, while those of Kamiokande-II require a delay of about 6.4~s. Establishing this unified, high-precision timeline is critical to test the physical nature of the first Kamiokande-II event (K1), specifically whether its distinct angular features signal the onset of the prompt neutronization burst. Previous literature often assumes  that K1 originated from inverse beta decay (IBD) on qualitative grounds. By contrast, our multidimensional likelihood analysis incorporates energy, direction, and MSW flavor-conversion effects to provide the first quantitative constraint on its origin. We find a likelihood ratio of 3--6 in favor of an accretion-phase $\bar{\nu}_e$ origin over a neutronization-burst origin, thereby turning a traditional assumption into a statistically bounded physical inference. This framework provides the most stringent constraints to date on the chronology of SN1987A and establishes a  precise methodology for future multimessenger observations of Galactic supernovae.
\end{abstract}

\maketitle

\newpage 
\newpage

\section{Introduction}
\label{sec:intro}
On 23 February 1987, three neutrino detectors (Kamiokande-II, IMB, Baksan) recorded a few dozen events
\cite{PhysRevLett.58.1490,PhysRevD.38.448,PhysRevLett.58.1494,PhysRevD.37.3361,ALEXEYEV1988209}
 which appeared to be produced in a common time interval.
These are the first neutrinos that can be attributed with certainty to an extrasolar source, the supernova SN1987A, and after so long they remain the only ones of this type.
They have been the focus of sustained investigations, as they represent a crucial diagnostic tool of the extreme environment of a supernova. \newline
Accurately reconstructing these explosion stages requires both precise
event timing and a reliable characterization of detector interactions.
If either is lacking, our ability to track the earliest phases of the
emission---such as the neutronization burst and accretion phase---is
significantly compromised. Precise timing provides valuable insights
into the neutrino flux, luminosity, and energy spectrum, whereas
mistimed events can heavily bias these estimates. Furthermore,
establishing the exact temporal sequence enables correlations with
other messengers, such as electromagnetic and gravitational-wave
signals, thereby strengthening a multimessenger approach.
In this work, we tackle both aspects by introducing a high-precision statistical framework; we demonstrate how precise timing calibrations and quantitative event-nature assessment rigorously constrain the chronology of the stellar collapse dynamics.

The temporal alignment of the events recorded by the three neutrino detectors has long remained an open issue, despite significant advances in neutrino physics, supernova dynamics, and statistical methods over the past two decades.
 While internal relative timing was robust, absolute UT offsets for Kamiokande-II and Baksan reached uncertainties of $\mathcal{O}$(1min), blurring the global collapse sequence~\cite{Fiorillo:2023frv}.
 We overcome these limitations by establishing a novel high-precision statistical framework.
 We introduce a rigorous mathematical distinction between Intrinsic
Timing Offsets (ITOs)---the stochastic delays inherent to discrete sampling in each detector---and
Relative Timing Offsets (RTOs), which quantify the absolute clock misalignments
between independent experiments. The ITOs are instrumental, detector-specific quantities, in principle affected by highly non-gaussian uncertainties and non-trivial correlations with the astrophysical parameters of the neutrino emission. They cannot, on their own, be naively used to establish whether an event in one detector precedes/follows an event in another one. We propose a new procedure using both offsets, in which the RTOs are directly extracted from data and a time reference is chosen, here the IMB clock. We also discuss corrections arising from differences in flight time across the Earth and from different measurements of SN1987A's celestial coordinates.

An important feature of our procedure is that it integrates the set of all measured quantities for each event---arrival time, energy, direction and background rate---while accounting for detector-specific response functions.
Incorporating all of this information is not merely an incremental improvement; it is essential for accurately reconstructing the timeline and thermodynamical properties of the explosion, moving beyond semi-qualitative analyses of the SN1987A data.

By constructing a newly defined $\chi^2$ function that explicitly minimizes over the RTOs and accounts for non-Gaussian statistical correlations, our method reduces the historical timing uncertainties by two orders of magnitude, reaching sub-second precision.
This framework effectively combines the three independent observations into a single, high-cadence record, enabling a more stringent reconstruction of the stellar-collapse dynamics.

We use a parametric model \cite{sym13101851,Bozza:2025wqo} to describe the temporal evolution of the $\bar{\nu}_e$ flux. We emphasize that our analysis
can also be readily repeated using a model derived from numerical simulations, with theoretical uncertainties incorporated where possible. Our results may then serve as a useful benchmark for comparison.
\newline
Another significant challenge concerns the nature of the detected events. Owing to its large cross section, inverse beta decay (IBD) $\bar\nu_e+p\to e^++n$ is typically assumed to be the only relevant neutrino interaction mode for the SN1987A dataset.
However, as already noted long ago \cite{Krivoruchenko:1988zg}, the angular distribution of IMB+Kamiokande-II is too forward-directed and does not fit well with the standard interpretation of IBD events.
To determine whether there is a way to ease tensions without invoking exotic physics, we must examine whether any event could have been caused by (anti)neutrino elastic scattering (ES)
${\nu}+e\to {\nu}+e$
\cite{Bahcall:1987ua,Krauss:1987re,Krivoruchenko:1988zg}.
The best ES candidate is K1, the first Kamiokande-II event \cite{Krivoruchenko:1988zg,Costantini:2004ry}, although the reconstructed angle of $(18\pm18)^{\circ}$
deviates slightly from
the direction of SN1987A and
 its measured energy, $20\pm 2.9\,\mbox{MeV}$, is
 quite high \cite{LoSecco:1988hb,Costantini:2004ry}.
 While previous literature estimated an ES probability of $\sim 30\%$ for
K1 based on purely qualitative or partial assessments under the accretion phase
alone \cite{Vissani:2014doa}, a comprehensive, multi-dimensional statistical constraint has been
lacking.
This remains a major unresolved issue when interpreting the SN1987A data as IBD events \cite{Bozza:2025wqo}, and it is a powerful motivation for our investigation.\\
An exciting possibility
involves again the time distribution:
K1 could be an ES event linked to
the early and intense neutronization burst emission and
due to the
 $p+e\to n+\nu_e$ processes
that occur at the onset of the collapse.
 Crucially, testing this scenario demands an
absolute chronology precise enough to resolve the sub-second transitions between
the prompt neutronization burst and the subsequent accretion phase. Moving beyond
traditional qualitative and partial assumptions \cite{LoSecco:1988hb}, we perform the first multi-dimensional
likelihood analysis that simultaneously accounts for neutrino arrival times, energy
spectra, non-Gaussian angular distributions, and MSW flavor oscillation scenarios
during the early collapse.
Based on conservative assumptions, we provide a
definitive quantitative bound on which neutrino emission phase K1 is most likely
to belong to, establishing a rigorous methodology for future multimessenger observations of Galactic supernovae.

Before proceeding with the analysis, we note that the issues addressed in this work represent core topics featured in prominent astroparticle, astrophysics, and neutrino physics textbooks. Crucial methodological challenges, including absolute timing uncertainties, the intrinsic ambiguity over the nature of the detected signals, and the complexity of background rejection, are widely covered from a pedagogical perspective in standard references such as \cite{giunti2007fundamentals,grupen2005astroparticle,xing2011neutrinos,rolfs1988cauldrons}. In this framework, the quantitative formalism and rigorous procedure introduced herein not only provide new results derived from the observed SN 1987A neutrino events, but also offer concrete tools to strengthen and enhance the educational treatment of these landmark events.

\section{Times of first neutrino event}\label{par:I}

In Table~\ref{Tabletimes}
we list the times $T_d^{\text{exp}}$ of the first detected neutrino event, as reported by each experiment~\cite{PhysRevLett.58.1494,PhysRevLett.58.1490,ALEXEYEV1988209}. The subscript $d=k,\,i,\,b$, here and in the following, stands for Kamiokande-II, IMB and Baksan, respectively.

The SN1987A detectors employed independent local timing systems lacking mutual coordination. While Kamiokande-II and Baksan reported uncertainties up to one minute, IMB achieved a superior absolute time measurement with a precision of 50~ms. We therefore adopt the IMB clock as the absolute reference for our global alignment. Under the hypothesis of a common neutrino source, we map the Kamiokande-II and Baksan events onto the IMB timeline to reconstruct a unified, absolute temporal sequence. Within this framework, the arrival time of the first detected $\bar{\nu}_e$ is denoted by $T_d$. By construction, the IMB absolute time $T_i$ coincides with
$T^{\text{exp}}_i$ yielding:
\begin{equation}
T_i = 7:35:41.374 \pm 0.050\,[\textbf{UT}].
\label{Tidaconfr}
\end{equation}

In this section, we proceed as follows: in subsection~\ref{par:I0} we present the adopted
parametrized model for the time-dependent $\bar{\nu}_e$ flux; in
subsection~\ref{par:Ia} we define the offset times; in subsection~\ref{par:Ib} we include the geometric effects due to the different positions of the three detectors; in subsection~\ref{par:Ic} we
calculate the absolute detection times of the first events in Kamiokande-II and Baksan, as well as the arrival time at Earth of the first neutrino from SN1987A.

\begin{table}[t]
\centering
\begin{tabular}{lc}
\toprule
Experiment & $T_d^{\mathrm{exp}}$ (\textbf{UT}) \\
\midrule
IMB & $7{:}35{:}41.374 \pm 0.050$ \\
Baksan & $7{:}36{:}11.818^{+2}_{-54}$ \\
Kamiokande-II & $7{:}35{:}35.000 \pm 60$ \\
\bottomrule
\end{tabular}
\caption{ Times of the first detected event in the UT standard format hh:mm:ss.sss~\cite{PhysRevLett.58.1494,PhysRevLett.58.1490,ALEXEYEV1988209}.}
\label{Tabletimes}
\end{table}

\subsection{The $\bar{\nu}_e$ flux}\label{par:I0}

We adopt the model of the time-dependent $\Bar{\nu}_e$ flux described in \cite{sym13101851}, that was successfully used in
\cite{Bozza:2025wqo}.\footnote{In this paper, we use the symbols $t$, $T$, and $\tau$ to denote times.
To avoid any confusion,
we use the symbol $\theta$ instead to denote temperature values.} This parametric function, unlike previous step-function models \cite{Loredo:2001rx,Pagliaroli:2008ur,Pagliaroli:2009qy,DedinNeto}, incorporates a continuous rise-time $t_{\mbox{\tiny max}}$ for a more robust global fit and a more precise scan of the temporal features of the emission. The model also contemplates a description both of the accretion and cooling phases.
In this model, the $\Bar{\nu}_e$ spectrum in the cooling phase follows a black-body radiation law
\begin{equation}
    \dv{\dot{N}_{\nu,c} }{E_{\nu}}(E_{\nu}, t) = \frac{c}{(hc)^3} \times \pi R^2_{ns} \times \frac{4 \pi E_{\nu}^2}{1 + \exp(E_{\nu}/\theta_c)} ,
\end{equation}
where $R_{ns}$ and $\theta_c$ are the radius and temperature of the nascent neutron star.
During the accretion phase, $\bar{\nu}_e$s are produced in the accreting region through the process
$e^++n\to p+\bar{\nu}_e$.
The corresponding spectrum is
\begin{equation}
    \frac{d\dot{N}_{\nu,a}}{dE_{\nu}} = \frac{c}{(hc)^3}  \times \frac{M_{\odot}}{m_n}\times \xi_n \times \frac{ 8 \pi E_e^2\,\sigma_{e^+ n}}{1 + \exp(E_e/\theta_a)} \,
\end{equation}
and depends on the neutron density $\xi_n$ and the temperature $\theta_a$ of the positron gas. The variation in time is obtained by
the assumption that $\xi_n$ and $\theta_c$ vary as an appropriate power of $\mathfrak{F}  (t)$, namely $\xi_n(t)=\xi_{n0}\times \mathfrak{F}(t)$ and $\theta_c(t)=\theta_{0}\times\sqrt[4]{\mathfrak{F}  (t)}$, where
\begin{equation}
 \mathfrak{F}  (t)=\left(\frac{1+\gamma\big(\frac{t_{\mbox{\tiny max}}}{\tau}\big)^{\gamma}}{\exp{\big[2\big(\frac{t}{\tau}\big)^\gamma-2\big(\frac{t_{\mbox{\tiny max}}}{\tau}\big)^{\gamma}\big]}+\gamma \big(\frac{t_{\mbox{\tiny max}}}{\tau}\big)^{\gamma}\big(\frac{t_{\mbox{\tiny max}}}{t}\big)^2}\right)^{\! \frac{1}{2}}\,.
 \label{eq:fcalligraphic}
\end{equation}
$\mathfrak{F}  (t)$ reproduces the temporal features of the luminosity as predicted by the simulations: an initial increasing phase up to the time $t_{\text{max}}$ (where $\mathfrak{F}  (t_{\text{max}})=1$), followed by a decreasing stage with different characteristic times $\tau_{(c,a)}$, for cooling and accretion, respectively. The shape parameter $\gamma$ is set to $1$ for the cooling and $2$ for the accretion.
The requirement that the average energy of neutrinos be a continuous function implies that $\theta_a = 0.6\, \theta_0$.

This model \cite{Bozza:2025wqo}
furnishes a long-sought realistic picture of the temporal evolution of the neutrino emission and permits to build a likelihood function that takes into account the measured times, energies, and scattering angles of SN1987A events.

 The astrophysical parameters of the model $R_{ns},\,\theta_0,\,\xi_{n0},\,\tau_a,$ and $\tau_c$, obtained from the fit procedure \cite{Bozza:2025wqo} are
 \begin{eqnarray}
      R_{\text{ns}0} &=& (17.0)^{+0.7}_{-0.5}\,\mbox{km},\quad  \xi_{n0} = (0.018)^{+0.025}_{-0.011}, \nonumber \\
 \theta_0 &=& (4.6)^{+0.5}_{-0.4}\,\mbox{MeV} \nonumber\\
 \tau_a &=& (0.52)^{+0.24}_{-0.15}\,  \text{s}, \quad \tau_c = (5.6)^{+1.8}_{-1.3} \,  \text{s}.\label{beobeo}
\end{eqnarray}
As discussed in \cite{Bozza:2025wqo}, the best fit model agrees well with expectations.
The small data set collected from SN1987A, implies a statistical error of the order of $1/\sqrt{N}\sim 20\%$ on the flux.

\subsection{Offset times}\label{par:Ia}

We formalize the temporal alignment of the SN1987A datasets by distinguishing between the {\em Intrinsic Timing Offset} (ITO) and the {\em Relative Timing Offset} (RTO).
\begin{itemize}
\item The ITOs (\(t_{i},t_{k},t_{b}\)) account for the stochastic delay inherent to discrete sampling of the initial neutrino flux at each of the three detectors (IMB, Kamiokande-II and Baksan) and are routinely estimated in contemporary emission models \cite{Loredo:2001rx,Pagliaroli:2008ur,Pagliaroli:2009qy,DedinNeto,Bozza:2025wqo}.
\item The RTOs (\(\Delta t_{k}\equiv t_{k}-t_{i}\) and \(\Delta t_{b}\equiv t_{b}-t_{i}\))
are the most convenient parameters
to quantify the timing misalignment between independent experiments, which is our goal.
\end{itemize}

It is worth stressing that this distinction is neither merely terminological nor equivalent to a trivial change of variables. For each detector, the ITO is, by definition, an internal quantity---the delay between the (unobserved) arrival of the first neutrino and the first detection of a neutrino event. As demonstrated in Figure 5 of \cite{Bozza:2025wqo}, the statistical distribution of the ITOs parameters---referred to as {\em delay times} in that work--- is strongly non-Gaussian. Consequently, ITOs alone cannot establish the absolute temporal
ordering of events recorded by independent, unsynchronized detectors.
A variable that relates each detector clock to a common reference is therefore required; this is precisely the role of the RTOs. In order to obtain the RTOs and reliably estimate their uncertainty intervals,
a rigorous, not-trivial, statistical procedure is required.

Marginalizing the likelihood over
the astrophysical parameters at $t_{\mbox{\tiny max}}= 100$~ms, we can derive $\mathcal{L}=\mathcal{L}(t_i,t_k,t_b)$; these parameters,
as shown in table 4 in \cite{Bozza:2025wqo}, are not correlated with the astrophysical parameters given in Eq.~\ref{beobeo}.
We obtain the values of the ITOs: $t_i = 43^{+102}_{-29}$~ms, $t_k = 35^{+65}_{-24}$~ms, $t_b = 54^{+152}_{-41}$~ms as in
\cite{Bozza:2025wqo}.

\begin{figure}[t!]
    \centering
    \subfloat[]{
        \includegraphics[width=0.45\textwidth, trim=0.94em 0em 1.88em 1em]{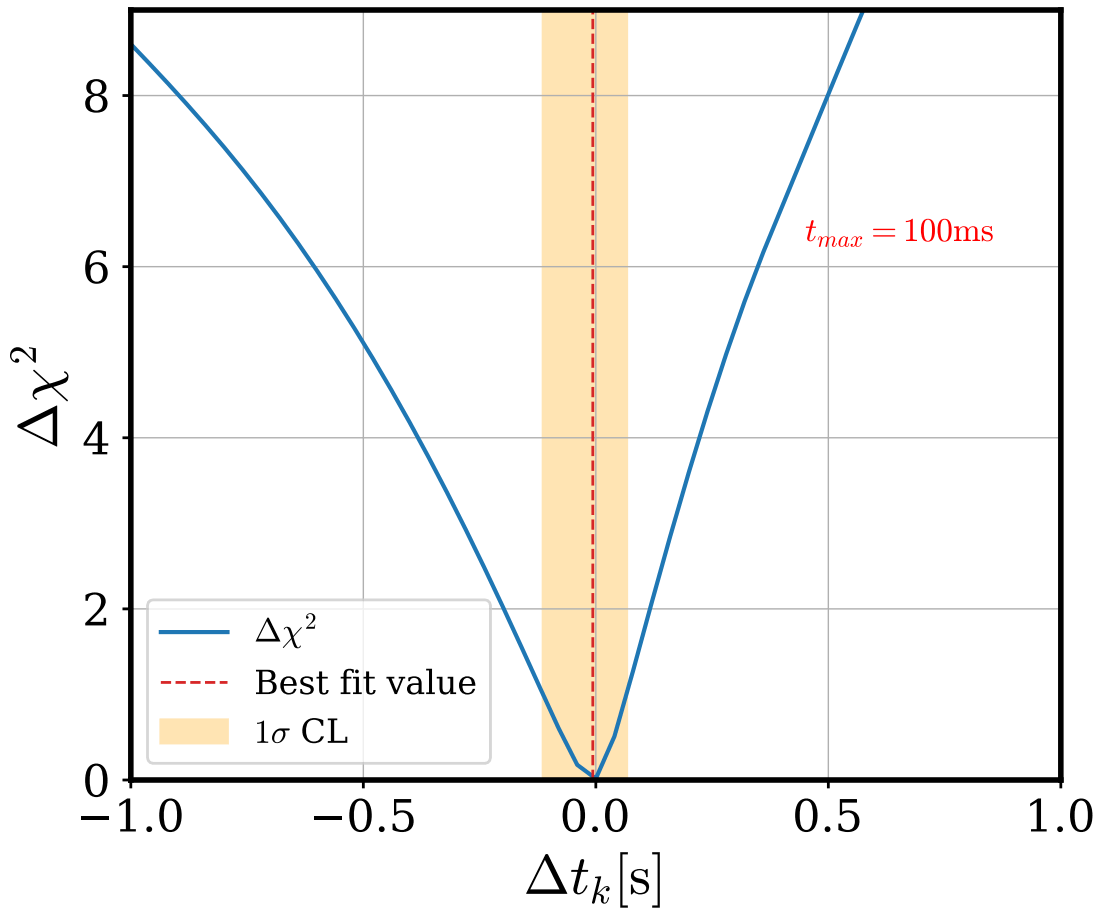}%
        \label{fig:profile_tk_ti}%
        }
    \hfill
    \subfloat[]{
        \includegraphics[width=0.45\textwidth, trim=2em 0em 0em 1em]{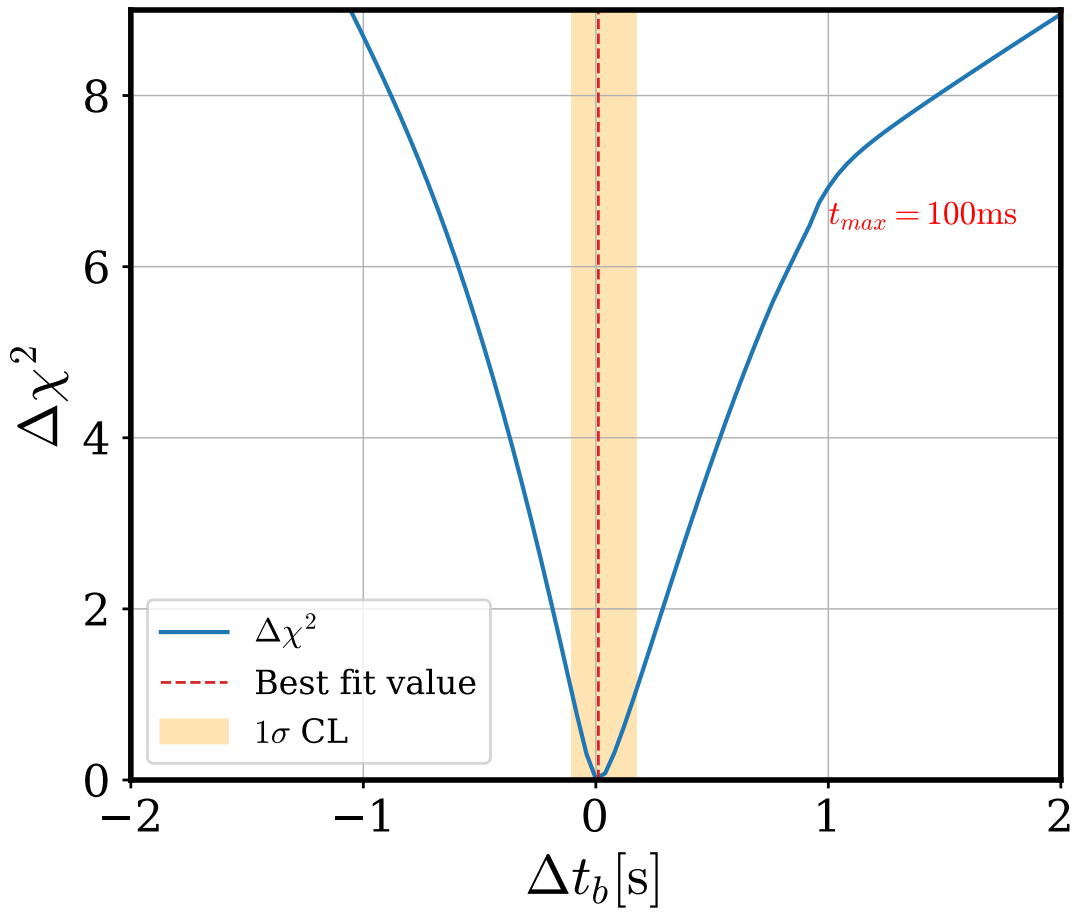}%
        \label{fig:profile_tb_ti}%
        }
    \caption{Profile likelihoods for the RTOs $\Delta t_{k}$ (left) and $\Delta t_{b}$ (right).
    The orange bands show the confidence interval at $1\sigma$,
    the red lines the best fit values.}
    \label{fig:time_diff_delay}
\end{figure}

\noindent
Next, we go beyond all the existing investigations of the SN1987A time profile, and, for the first time, obtain statistically significant values of the RTOs.
 We achieve this by defining a new $\chi^2-$function that depends explicitly on the RTOs and the ITO of IMB:
\begin{equation}
    \chi^2(t_i,\Delta t_k,\Delta t_b)\equiv-2\log\mathcal{L}(t_i,t_i+\Delta t_k, t_i+\Delta t_b).
\end{equation}
The IMB clock is taken as the reference.
The minimization of the newly defined $\chi^2-$function allows us to determine the RTOs directly.
 We find the best-fit values (see Fig.~\ref{fig:time_diff_delay}):
 \begin{equation}\label{eq:rto}
 \Delta t_k=-7^{+73}_{-107}\,\mbox{ms},\quad  \Delta t_b=10^{+157}_{-112}\,\mbox{ms}.
 \end{equation}
 These results are new.
 A complete comparison with the results of \cite{Abbott:1987bm} is not possible, since their Table~2 reports only the best-fit values of $\Delta t_k$ for two nonzero neutrino masses, $m=1$ and $3$ eV. Nevertheless, the results obtained for $m=1$ eV using their ``hot ball'' astrophysical model are consistent, within uncertainties, with those reported in Eq.~\ref{eq:rto}.

While a thorough comparison with the previous literature will be presented in section \ref{Comparison with  previous literature}, we emphasize here only that both the error bars and the Baksan RTO have been estimated for the first time.

Let us denote by
$\bar{T}_0$ the time at which the first neutrino (not detected) arrives at the IMB detector.
Including the uncertainty of $50\,\mbox{ms}$ in $T_{i}$, we find
 \begin{equation}
     \bar{T}_0= T_i- t_i=7:35:41.292^{+0.072}_{-0.117}\,[\textbf{UT}].
\end{equation}
The procedure for obtaining $\bar{T}_0$ is summarized in ~\ref{AppendixA}.

In the limit where the Earth's dimensions are neglected, all detectors can be treated as co-located with IMB, implying a common value of
$\bar{T}_0$. That implies the equality $\bar{T}_{k(b)}-t_{k(b)}=T_i-t_i$, which yields
\begin{equation}
    \Bar{T}_{k(b)}
    =
    T_{i}+\Delta t_{k(b)},
\end{equation}

 \noindent where we use the best-fit values of $\Delta t_k$ and $\Delta t_b$.

 \begin{table*}[t]
\centering
\begin{tabular}{l c c c c}
\toprule
$t_{\text{max}} (\mbox{s})$ & 0.050 & 0.100 & 0.150 & 0.200 \\
\midrule
$\Delta t_k (\mbox{s})$ & $-0.002^{+0.056}_{-0.092}$ & $-0.007^{+0.073}_{-0.107}$ & $-0.117^{+0.090}_{-0.122}$ & $-0.015^{+0.107}_{-0.139}$ \\
\addlinespace
$\Delta t_b (\mbox{s})$ & $0.002^{+0.143}_{-0.092}$ & $0.010^{+0.157}_{-0.112}$ & $0.021^{+0.171}_{-0.134}$ & $0.031^{+0.184}_{-0.156}$ \\
\bottomrule
\end{tabular}
\caption{RTO values and uncertainties as a function of $t_{\text{max}}$.}
\label{tab:value_tt}
\end{table*}

  As the fiducial value of $t_{\text{max}}$ could in principle
affect the previous results, we have repeated the minimization of $\chi^2(t_i,\Delta t_k, \Delta t_b)$ by varying $t_{\text{max}}$,
obtaining the quantitative estimate of the impact of the rising phase in the reconstruction of SN1987A time-line.
 We find that the dependence of the RTOs on the specific value of $t_{\mbox{\tiny max}}$ is mild across a wide, physically motivated range. Their distributions only exhibit minor variations, of the order of $10^{-2}$ s, which are consistent within the uncertainties and indicate no statistically significant
discrepancy. See Table~\ref{tab:value_tt} for the numerical values of the RTOs corresponding to different rising times. We see that varying the parameter $t_{\text{max}}$ from its reference value results in only a modest change in uncertainties. We can cover the entire range of plausible values shown in the table simply by extending the uncertainty ranges of the parameter $\Delta t_k$ by $+34$ ms and $-32$ ms and those of the parameter $\Delta t_b$ by $+27$ ms and $-44$ ms, corresponding to an overall variation of about $30\,\%$ with respect to the reference values for $t_{\text{max}}=100\,\text{ms}$.

\subsection{Transit times}\label{par:Ib}
The geographic separation of the detectors necessitates accounting for the neutrino wave-front transit times, $\delta t_d$, relative to the IMB site. Given the Southern sky origin of SN1987A, the neutrino flux propagated through the Earth to reach the Northern hemisphere detectors. The geometric path lengths are determined by $L_d = -2\mathcal{R} \langle\vec{n}_*, \vec{n}_d\rangle$, where $\mathcal{R}$ is the Earth's radius, $\vec{n}_d$ is the detector's geocentric unit vector, and $\vec{n}_*$ is the unit vector toward the supernova at $T_i$ (computed using
\href{https://aa.usno.navy.mil/data/siderealtime}{USNO Sidereal Time}). Using detector coordinates from \cite{1970CoTol..89.....S,Hosaka2006,Gajewski:1990kv,Kuzminov2012} (see Table~\ref{Table:position}), we calculate the relative transit times:
\begin{equation}
\delta t_{k} = \frac{L_i-L_k}{2 c} \simeq 7\,\text{ms}, \quad
\delta t_{b} = \frac{L_i-L_b}{2 c} \simeq -3\,\text{ms}.
\end{equation}

The procedure for extracting the transit times from the information on the geographic positions of the detectors and celestial coordinates of the star is described in detail in \,\ref{AppendixB}.
Different measurements of the SN1987A position are considered \cite{Bouchet:2024ivj,2012PASP..124..668Y} and we show that no significant discrepancies arise in the result.
Consequently, the wave-front reached Kamiokande-II first, followed by IMB, and finally Baksan.
Our results for $L_d$ agree with those in~\cite{LunardiniSmirnov2001} within $1\%$.

Using $T_k$ and $T_b$ rather than $T_i$ to calculate $\vec{n}_*$, does not affect the calculation, as the effects of the rotation of Earth are completely negligible.
These sub-10 ms corrections are incorporated into our global alignment but are negligible compared to the 50 ms IMB absolute timing uncertainty.

\begin{table*}[t]
\centering
\begin{tabular}{l c c c c}
\toprule
Star & Declination & Time ($T_i$) & RA & $D$ \\
\midrule
Sanduleak -69$^\circ$202 & $-69^\circ 16'11.07''$ & $7^{\mathrm{h}}\,35^{\mathrm{m}}\,41.374^{\mathrm{s}}$ & $5^{\mathrm{h}}\,35^{\mathrm{m}}\,27.92^{\mathrm{s}}$ & 51.4 kpc \\
\bottomrule\\[-0.4ex]
\toprule
Detector & Latitude & Longitude & LST & $L_d$ \\
\midrule
IMB & $41.756^\circ$ & $-81.286^\circ$ & $12^{\mathrm{h}}\,21^{\mathrm{m}}\,12.24^{\mathrm{s}}$ & 8600 km \\
Kamiokande-II & $36.4225^\circ$ & $137.3153^\circ$ & $2^{\mathrm{h}}\,55^{\mathrm{m}}\,36.55^{\mathrm{s}}$ & 4300 km \\
Baksan & $43^\circ\,16'\,32''$ & $42^\circ\,41'\,25''$ & $20^{\mathrm{h}}\,37^{\mathrm{m}}\,6.55^{\mathrm{s}}$ & 10500 km \\
\bottomrule
\end{tabular}
\caption{ Position data for the exploding supernova \cite{1970CoTol..89.....S} and the detectors \cite{Hosaka2006,Gajewski:1990kv,Kuzminov2012}. The LST of each detector is calculated at the time $T_i$ using \href{https://aa.usno.navy.mil/dat/siderealtime}{USNO Sidereal Time}.}
\label{Table:position}
\end{table*}

\subsection{Absolute times}\label{par:Ic}
The first neutrino arrived on Earth at the time
$T_0=\bar{T}_{0}- \delta T_{i}$,
where  $\delta T_{i}= (\mathcal{R}+L_i/2)/{c}=36\mbox{ ms}$,
that is a little bit earlier than its arrival at IMB.
Thus, the Earth was reached by SN1987A neutrinos at \begin{equation}T_0=7:35:41.256^{+0.072}_{-0.117}\, [\textbf{UT}].
\end{equation}
The symbols
$T_k= \Bar{T}_k-\delta t_k$ and
      $T_b= \Bar{T}_b-\delta t_b$ indicate the absolute times
of the first $\bar{\nu}_e$ event detected in the Kamiokande-II and Baksan locations, including corrections for finite transit times.
Following the procedure detailed in~\ref{AppendixA}, we finally obtain
\begin{eqnarray}
      T_k &=& 7:35:41.350_{-0.129}^{ +0.106}\,[\textbf{UT}]\,,
     \\         T_b &=& 7:35:41.392_{-0.140}^{+0.167}\,[\textbf{UT}]\,,
\end{eqnarray}
to be compared with $T_d^{\text{exp}}$ in Table~\ref{Tabletimes}.
For Kamiokande-II, we need to add 6.350~s to the quoted absolute times; for Baksan, we need to subtract 30.426~s from the quoted absolute times. In both cases, the correction is compatible with the large uncertainties of the absolute times.

A conservative procedure for estimating the effects of changes in $t_{\text{max}}$ is to widen the regions of uncertainty of the RTOs without changing the central values, as described at the end of section \ref{par:Ia}. This procedure yields the following values: $T_k=7:35:41:345^{+0.129}_{-0.153}$, and $T_b=7:35:41:412^{+0.197}_{-0.173}$; as we can see, these changes are not essential to our results.

The impact of astrophysical uncertainties has been assessed in Ref.~\cite{Bozza:2025wqo}, where the corresponding correlation matrix is reported in Table~4. The ITOs exhibit negligible correlations with the astrophysical parameters, implying that their effect on the synchronization procedure is subleading compared to the uncertainty associated with the choice of $t_{\text{max}}$. Likewise, flavour-conversion effects are expected to play only a minor role in the currently favoured normal mass-ordering scenario\,\cite{Bozza:2025wqo}: in April 2026, the preference for the spectrum with a “normal” ordering  was quantified to $\Delta \chi^2=9.4$ (see Ref.\,\cite{concha2026}), i.e., about 3 Gaussian sigmas. The inclusion of the 207.2 days of JUNO data in the preliminary fits of \cite{Marcone:NOW2026,Lisi:2026SIF}  increased the value of sigmas to 4. This constitutes a significant preference, based purely on empirical evidence, for the type of spectrum that deviates the least from expectations—even though current theoretical  expectations are not prescriptive in nature.
In the inverted-ordering case, their impact remains limited, affecting primarily the accretion phase of the neutrino signal \cite{DedinNeto:2023hhp}.

We also acknowledge that the synchronization may depend to some extent on the modelling of the earliest rising phase of the neutrino signal. Alternative descriptions of the initial neutrino emission could result in a slightly different determination of the synchronization time. However, based on the typical duration and variability of the initial rising phase, we expect such effects to remain limited and not to exceed the sub-second level.

\begin{figure*}[t!]
  \centering
\subfloat{\includegraphics[width=1\linewidth, trim=-0.0em 0em 4em 5em]{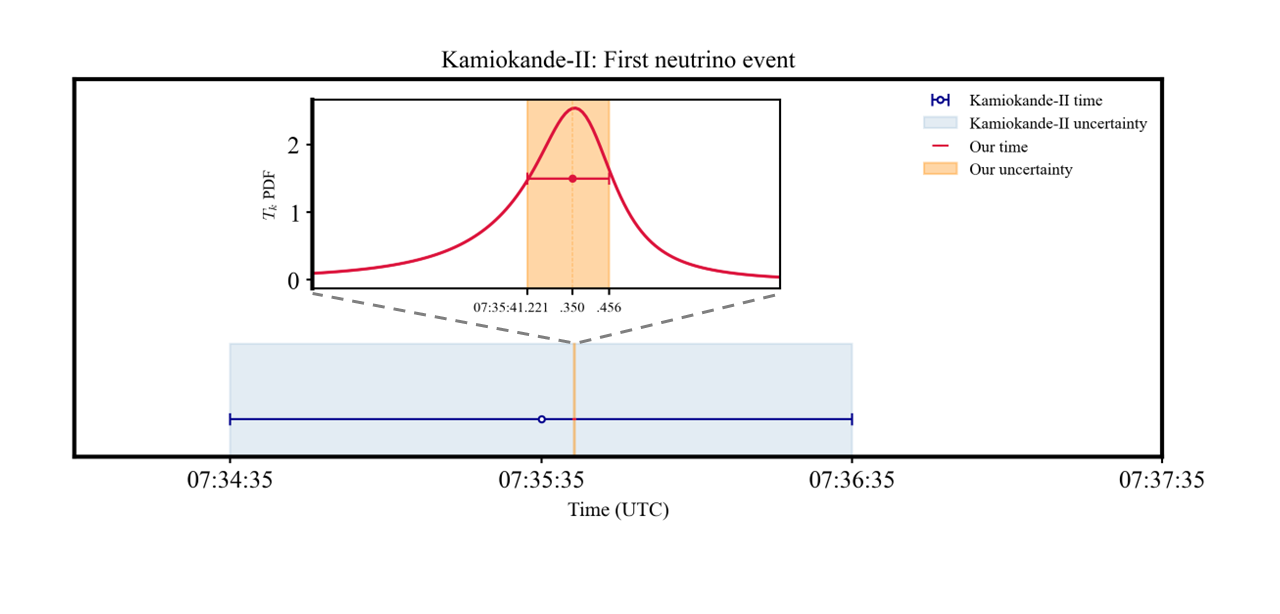}\label{fig:T_k}}
  \vspace{-0.054\textwidth}
  \centering
\subfloat{\includegraphics[width=.95\linewidth, trim=-.15in .0in .0em .0in]{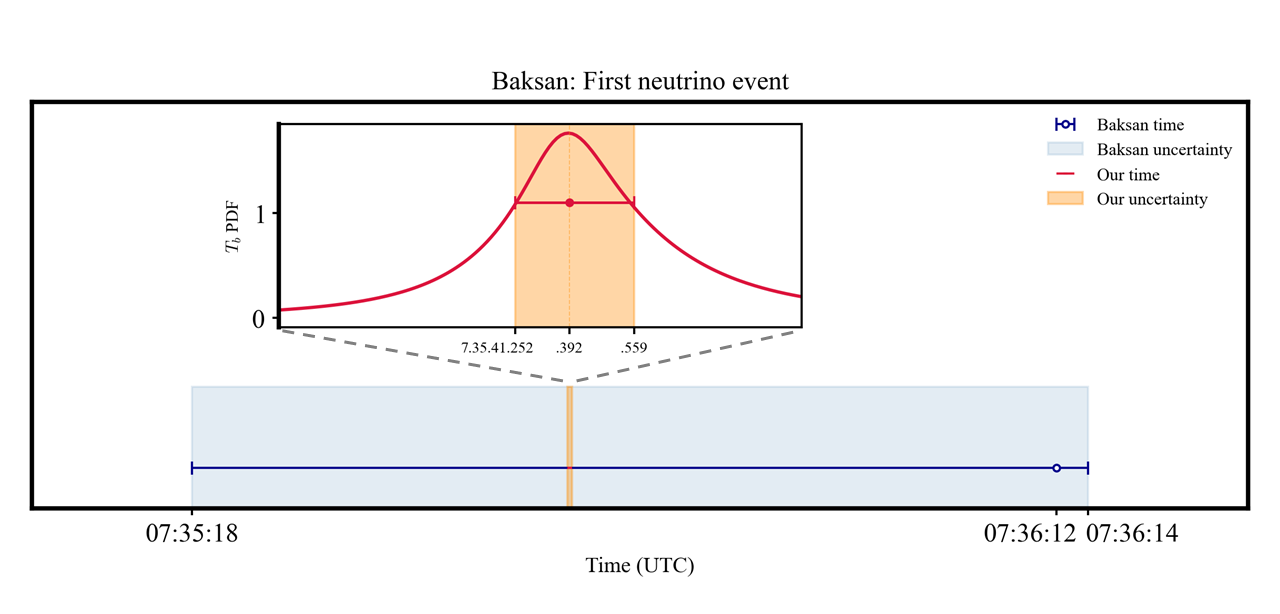}\label{fig:T_b}}
   \vskip-4mm
  \caption{Comparison of the measured times $T^{\text{exp}}_k$ and $T^{\text{exp}}_b$ (light blue bands) with our
  estimates $T_k$ and $T_b$.
  The inset shows the probability density function and
  our uncertainty estimate (orange bands).}
  \label{fig:times_comparison}
\end{figure*}

The uncertainties on the time of first detection are reduced a posteriori by {\em two} orders of magnitude with respect to the estimates provided by the clocks of the two experiments.
This is visualized in Fig.~\ref{fig:times_comparison} through the relative size of the blue band (experimental uncertainties) and the orange band (uncertainties after the alignment procedure).
The central values suggest that the first observed event was Kamiokande-II's event K1.
These results and the estimated uncertainties are new.

\section{The elastic scattering hypothesis}

\label{sec:es_hypothesis}
The event K1 is of particular interest. It is the most suitable event to originate from ES rather than IBD \cite{Krivoruchenko:1988zg,Costantini:2004ry}.
Rather than excluding the possibility of ES events by just looking at the raw and apparently big values of scattering angles in SN1987A dataset,
we extend the discussions in \cite{Krivoruchenko:1988zg,Costantini:2004ry,Vissani:2014doa} by presenting a novel quantitative $\chi^2$ analysis of K1 that incorporates the full multidimensional information contained in the data, together with a refined description of the detector's angular response.

The preceding discussion of the absolute timing indicates that K1 was likely the earliest of all the recorded events,
 thereby placing on well-motivated statistical grounds a further intriguing possibility that has occasionally been raised in the literature:
 that K1 was an ES event caused by a {\em neutrino} produced before the accretion and cooling phases, during the neutronization phase (NP).
Thus, K1 could correspond to the first (and only) supernova neutrino ever detected.
 If K1 originated from NP, its temporal location
would be more precise, as the emission rises quickly and lasts around 10 ms only. Given the robustness of the relative times, the timing accuracy of all Kamiokande-II events would further improve.

Given its potentially significant impact on the reconstruction of the SN1987A timeline and on our understanding of the early stages of the explosion, we now evaluate this hypothesis quantitatively against the standard interpretation of K1 as an IBD event. A reliable comparison between the ES and IBD hypotheses requires an accurate treatment of the energy and angular responses of Kamiokande-II. In particular, the ES event rate must be convolved with the angular response given in Eq.~(31) of \cite{Ianni:2009bd} to obtain the distribution of the observed direction.

The analysis developed for K1 lays the foundation for a quantitative method to distinguish ES from IBD events and to infer their origin in the much larger neutrino data sample expected from the next Galactic core-collapse supernova. In particular, it allows to assign a probability of being an ES event to each detection, moving beyond naive classifications based only on partial assumptions and on the nominal values of the energy, arrival time, and scattering angle.

It is presented as follows. In subsection~\ref{subsec:neutronizationphaseflux}, we describe the neutrino-flux model during the NP. In subsection~\ref{subsec:angularresponse}, we discuss the response of Kamiokande-II, with particular emphasis on its angular resolution and the effects of neutrino oscillations. Finally, in subsection~\ref{subsec:results}, we discuss our results.

\subsection{The Neutronization Phase flux}
\label{subsec:neutronizationphaseflux}
During NP, the collapsing core undergoes a
rapid neutronization through electron capture processes $p + e^- \rightarrow n + \nu_e,$
that lead to a sudden burst of electron neutrinos, the so-called \emph{neutronization burst}, which represents the first observable neutrino signal from a core-collapse supernova.
In order to estimate the probability that the first detected event originates from ES during the NP,
we need to consider a model for the neutrino flux.

Unlike later phases of the explosion, the neutrino spectrum during the NP is expected to deviate from a purely thermal (Fermi--Dirac) distribution and the energy spectrum can be effectively modeled by a modified Maxwell--Boltzmann distribution
\cite{Keil_2003}
\begin{equation}\label{pinch}
    \varphi(E_\nu) = \frac{E_\nu^\alpha \, e^{-E_\nu/\Theta}}{\Theta^{1+\alpha} \, \Gamma(1+\alpha)}
    \mbox{ with }
 \Theta = \frac{\bar{\epsilon}}{1 + \alpha},
 \end{equation}
 where $\bar{\epsilon}$ is the mean neutrino energy. This parameterization accounts for the spectral ``pinching'' caused by neutrino transport and opacity effects in the dense stellar medium via the parameter $\alpha$; typically, $\alpha = 5$ during the NP. The corresponding electron neutrino flux peak at Earth is estimated from the maximum luminosity $\mathcal{L}_{\text{max}}$:
\begin{equation}
    \Phi_{\text{max}} = \frac{1}{4\pi D^2}
    \frac{\mathcal{L}_{\text{max}}}{\bar{\epsilon}},
\end{equation}
where $D = 51.4\,\mbox{kpc}$ is the distance to SN1987A. We adopt the values
\begin{equation}\label{prm}
\mathcal{L}_{\text{max}} = (4 \pm 2) \times 10^{53}\,{\mbox{erg}}/{\mbox{s}},\
\bar{\epsilon}=13\,\mbox{MeV},\
\alpha=5,
\end{equation}
consistent with Figure~9 of \cite{Mirizzi:2015eza} and corresponding to $\Phi_{\text{max}}= 6.1 \times 10^{10} \, \frac{\nu_e}{\text{cm}^2\,\text{s}}$.
The value of the peak luminosity $\mathcal{L}_{\text{max}}$ is important, as the ES cross section is proportional to the neutrino energy \cite{Costantini:2004ry}.
The choice $\bar{\epsilon}=13,\mbox{MeV}$, which is close to the maximum average energy expected during the neutronization phase, ensures that the likelihood of K1 being an ES event is not underestimated.
The time-energy dependent flux at Earth can be expressed as
\begin{equation}
    \frac{ d \Phi}{dE_\nu}(E_\nu, t)
    = \Phi_{\text{max}} \times \varphi(E_\nu)\times\, \mathfrak{F}(t; t_{\text{max}}, \tau, \gamma)\,,
\end{equation}
following a parameterization similar to that proposed in \cite{sym13101851,Bozza:2025wqo}. The $\mathfrak{F}$, given in Eq.\,\eqref{eq:fcalligraphic},
encodes the temporal evolution of the burst, peaked at $t_{\text{max}}$, with characteristic duration $\tau$ and shape parameters $\gamma$.
In order to reproduce the temporal shape of the flux indicated by simulations, we set $t_{\text{max}}=2.5\,\mbox{ms},\,\tau=8.2\,\mbox{ms},\,\gamma=2$.
This implies that
$\Delta t=\int \mathfrak{F}(t_{\text{max}}; t_{\text{max}}, \tau, \gamma)\, dt=7.5\,\mbox{ms}$; thus, the energy radiated during the NP is $\mathcal{E}=
\Delta t \times \mathcal{L}_{\text{max}}=
3 \times 10^{51} \, \mbox{erg}$, which is in the right ballpark.
The flux emitted during the neutronization phase is known with reasonable accuracy; the description we have just presented ensures that we do not understimate its effect with regard the occurrence of ES events.

\subsection{The ES signal and detector response}
\label{subsec:angularresponse}
The differential ES cross section for electron-neutrino scattering
can be expressed as
\begin{equation}
\frac{d\sigma_{\nu_{(e,\mu,\tau)}}^{\text{ES}}}{dT_e} = \frac{2 G_F^2 m_e}{\pi}
\left[
    g_L^2
    + g_R^2 \left(1 - \frac{T_e}{E_\nu}\right)^2
    - g_L g_R \frac{m_e T_e}{E_\nu^2}
\right],
\end{equation}
where $g_L$, $g_R$ are the left- and right-handed coupling constants,
whose values depend on the neutrino flavor. The probability of detecting an ES signal during NP is given by the double-differential event rate
\begin{equation}
\label{formula:pp}
\begin{split}
S_e(T_e, \cos{\theta})=&\frac{d^2N^{\text{ES}}}{dT_e \, d\cos\theta}(T_e,\cos\theta)=\\
&= N_e\,  \frac{d \Phi}{dE_\nu}(E_\nu, t)\left| \frac{\partial E_\nu}{\partial \cos\theta}\right| \times \bigg(P_{ee}
\frac{d\sigma_{\nu_e}^{\text{ES}}}{dT_e}
+(1-P_{ee})
\frac{d\sigma_{\nu_{(\mu,\tau)}}^{\text{ES}}}{dT_e}
\bigg)
\end{split}
\end{equation}
The first contribution is proportional to the survival probability of electron neutrinos $P_{ee}$ and refers to ES events from electronic neutrinos, whereas the second contribution refers to ES events from muonic and tauonic neutrinos generated through flavor conversion.
In Eq.\,\eqref{formula:pp}, $\cos\theta =  \vec{n} \cdot \vec{n}^*$ ($\vec{n}$ is the direction of emission of the electron),
$T_e$ is the kinetic energy of the recoiling electron and
$E_\nu=E_\nu(T_e, \cos\theta)$ is the incoming neutrino energy.
The observed signal is modulated by the detector resolution in both energy and angle.
The energy response function can be found in~\cite{Bozza:2025wqo} and is given by the product of the intrinsic detector efficiency $\eta(E_e)$ and the energy-resolution function $G(E_e-E_j,\sigma(E_e))$. For the angular response we adopt the formulation of eqs.~(31-33) in~\cite{Ianni:2009bd}:
\begin{equation}
\frac{d\rho}{d\mu} = \frac{1}{N_2\,\delta n_i^2}
\exp\!\left[-\frac{\sqrt{2(1 - \mu)}}{\delta n_i}\right].
\label{eq:ang_response}
\end{equation}
We define the reconstructed electron direction $\vec{n}_i$ and indicate as $\phi$ the azimuthal angle around it. Then
$\mu = \vec{n}\cdot\vec{n}_i$,
$N_2 = 1 - (1 + 2/\delta n_i)e^{-2/\delta n_i}$ to ensure normalization, and
$\delta n_i \simeq 0.424\times  \delta \theta_i \, (1 - \delta\theta_i^2/24)$, where $\theta_i$ is the angle between the directions $\vec{n}$ and $\vec{n}_i$. The ES differential signal depends on $\vec{n}\cdot\vec{n}^*$, which is given by $\vec{n}\cdot\vec{n}^*=\cos{\theta_i}\cos{\theta}+\sin{\theta_i}\sin{\theta}\cos{\phi}$.

Given the observed energy $E_j$ and the emission direction $\vec{n}_j$, the expected rate of the ES signal $S(t)$ can be calculated by integrating
the true rate $S_e$, weighted with the energy and the angular response functions,
over $E_e=T_e+m_e$ and the possible directions of
  $\vec{n}$~\cite{Ianni:2009bd}.
The full likelihood function related to ES processes is given by
\begin{equation}
\begin{split}
\mathcal{L}_d&=\prod_{j}^{N_d}e^{S(t_d+\delta t_j+\tau_d/2)\,\tau_d}\times
 \Bigg[\frac{B_j}{2}+\\ & +\int dE_e\,\eta(E_e)\,G(E_e,E_j,\sigma(E_e)) \int\frac{d\phi\, d\mu}{2\pi}\\ & \frac{d\rho}{d\mu}(\mu,\delta n_j)\,
S_e(E_e,\vec{n}\cdot\vec{n}^{\,*},
\Delta t_k+t_i+\delta t_j)
\Bigg]
\end{split}
\end{equation}
where $j$ denotes the $j$-th event detected by Kamiokande-II, $B_j$ its differential background and $\delta t_j$ the time relative to the first detection.
The computation of the likelihood function is highly demanding because it involves a three-dimensional integration over energy and angles, together with the nontrivial shapes of the response functions.

Focusing on K1, we have $\delta\theta_i=18^\circ$.
This procedure allows us to account for the directional reconstruction uncertainty, introducing non-Gaussian tails that provide a more realistic description of the detector performance than a simple Gaussian model.
It provides the most accurate estimate of the probability that the detected event is due to ES, overcoming the qualitative or incomplete assumptions at the root of previous calculations.

We evaluate the effects of flavor conversion using MSW theory \cite{Dighe:1999bi} and the measured mixing parameters. While supernova neutrino oscillations are generally known to be complicated, it is noticeable that during the NP a very remarkable simplification occurs.
Citing \cite{Volpe:2023met}:
``From the point of view of flavor evolution, the neutronization-burst represents a unique phase. Only the MSW effect appears to influence the neutrino spectra''.
Therefore we expect a full conversion to mass eigenstates, $\nu_e\to \nu_3$ ($\nu_e\to \nu_2$) for normal (inverted) ordering.
  The survival probability of electron neutrinos is
\begin{equation}
P_{ee}=\left\{
\begin{array}{cc}
\sin^2\theta_{13}=0.022  & \mbox{for normal ordering}\\[1ex] \cos^2{\theta_{13}}\sin^2{\theta_{12}}=0.296 & \mbox{for inverted ordering.}
\end{array}\right.
\end{equation}
Thus, the initial flavor changes significantly from $\nu_e$ to $\nu_{\mu/\tau}$, decreasing the ES interaction rate.

  \subsection{Results and discussion}
 \label{subsec:results}
\begin{figure*}[t]
\includegraphics[width=0.9\textwidth]{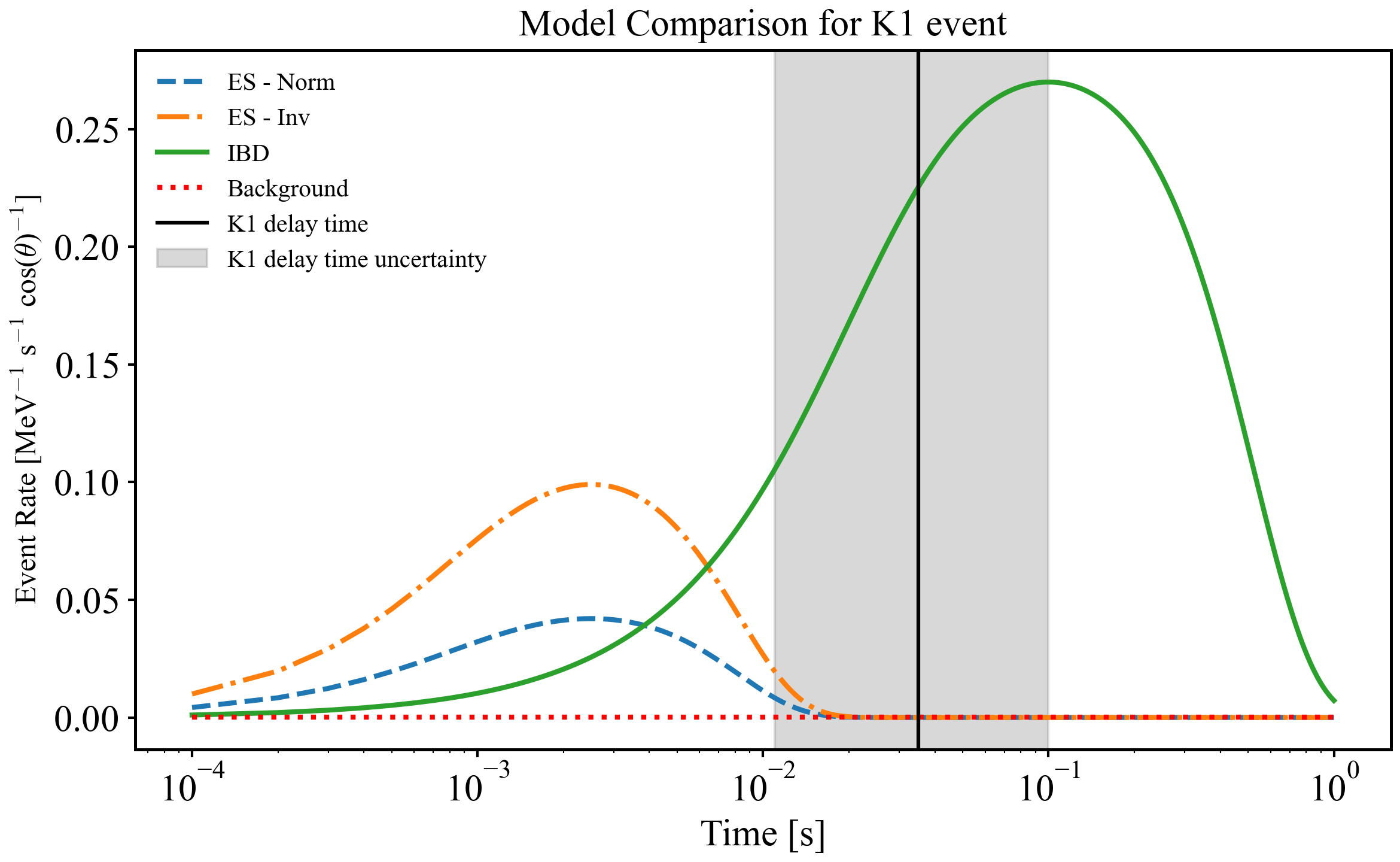}

\caption{\small Comparison of the expected event rates for
IBD (solid), ES (dashed for normal ordering, dashed dot for inverted ordering) and background (dotted red) as a function of logarithmically scaled time.
The offset time of the K1 event from the fit and the estimated uncertainties are indicated by the vertical black line and the shaded region.
}
\label{fig:K1comparison}
\end{figure*}

The temporal profile of the expected event rates, shown in Fig.~\ref{fig:K1comparison},
summarizes our discussion. Based on the measured energy and angular information of the event K1, we observe that the likelihood function possesses two local maxima, the greatest one corresponding to the IBD scenario. We give a conservative upper bound on the ES hypothesis considering the maximum likelihood of an ES origin during the NP (dashed orange and blu curves) and obtain $0.042$ for normal mass ordering and $0.10$ for inverted mass ordering according to the MSW theory~\cite{Dighe:1999bi}. The corresponding likelihood for the IBD hypothesis (solide green curve) is found to be $0.27$ while the background is negligible. 

Although there is a non-negligible likelihood that the K1 event has an ES origin, the values of the likelihoods indicate that the IBD channel is 3--6 times more likely, for the inverted/normal ordering, respectively.
The ES contribution remains subdominant, especially when the mass spectrum is normal, since the survival probability $P_{ee}$ is small
due to the adiabatic conversion to the mass eigenstate $\nu_e\rightarrow \nu_3$~\cite{Dighe:1999bi}
and the cross section of $\nu_\mu$ and
$\nu_\tau$ is six times smaller than that of
 $\nu_e$.
The likelihood ratio $3-6$ puts a mere upper bound to the likelihood of the hypothesis of an ES from a NP originated event. Only a radical change in the NP model (or perhaps some independent information about absolute times--which is currently purely speculative) could alter the conclusion: for instance, if the NP had larger
$\mathcal{L}_{\text{max}}$,
$\bar{\epsilon}$, if speculative (and at present unexpected) new oscillation effects would produce $P_{ee}\sim 1$ and/or if the exact time of the onset of the explosion is measured. Recall that the normal mass ordering is significantly preferred by current data analyses, see section \,\ref{par:Ic}.

Based on the temporal distribution of the events, our results show that the maximum-likelihood hypothesis of K1 being an ES event from a neutrino emitted during the NP is poorly significant compared to the standard IBD interpretation. Together with the result shown in Figure 2 of \cite{Vissani:2008wbj}, where the accretion+cooling neutrino fluence (\emph{i.e.} time-integrated flux) is used to test the ES hypothesis, we conclude that the SN1987A data do not support the presence of ES events. The anomalies in the angular distribution were likely produced either by a statistical fluctuation or by unknown systematic uncertainties. The possibility that some events originated from Beyond Standard Model physics cannot be excluded.

It is interesting to note that if the timing of the NP phase were known, the maximum-likelihood ratio could be computed at the exact detection time. In the event of a future core-collapse supernova, the detection of gravitational waves from the collapse will allow the onset of NP to be constrained with millisecond precision, opening the possibility of investigating the ES hypothesis with much greater precision than the upper-bound estimate proposed here.

\section{Comparison with previous literature}
\label{Comparison with  previous literature}
To clearly contextualize the methodological evolution of SN1987A data analysis, Table~\ref{tab:lit_comparison_final}, based on \cite{Bahcall:1987ua,Arafune:1987ua,Sato:1987rd,Abbott:1987bm,Chudakov:1988qb,LoSecco:1988hb,Krivoruchenko:1988zg,Loredo:1988mk,Kielczewska:1990zf,Smirnov,kernan1994yet,Jegerlehner:1996kx,Loredo:2001rx,Costantini:2004ry,Vissani:2008wbj,Pagliaroli:2008ur,Fiorillo:2023frv,DedinNeto,Ivanez-Ballesteros:2023lqa,Bozza:2025wqo,Burrows:1987zz,Spergel:1987ex}, tracks how individual physical and statistical elements were progressively incorporated over the last four decades.
\vspace{0.4cm}
\begin{table*}[ht]
\centering

\begin{tabular}{lcccccc}
\toprule
\textbf{Authors} & \textbf{Offset} & \textbf{ES} & \textbf{Energy} & \textbf{Time} & \textbf{Angle} & \textbf{$\nu$-osc} \\
\midrule

\href{https://doi.org/10.1038/327682a0}{Bahcall et al.\ 87}
  & ---          & \yes& \no           & \no           & \no           & \yes\\

\href{https://doi.org/10.1103/PhysRevLett.59.367}{Arafune \& Fukugita 87}
  & ---          & \yes          & \no           & \no           & \no           & \yes          \\

\href{https://doi.org/10.1103/PhysRevLett.58.2722}{Sato \& Suzuki 87}
  & ---          & \yes          & \no           & \no           & \no           & \no           \\

  \href{https://ui.adsabs.harvard.edu/link_gateway/1987ApJ...318L..63B/doi:10.1086/184938}{Burrows et al. 87}
    & {RTO} & \no           &  \yes& \yes& \no           & \no           \\

\href{https://doi.org/10.1016/0550-3213(88)90371-9}{Abbott et al.\ 88}
  &  {RTO} & \no           &  \yes&  \yes& \no           & \no           \\

\href{https://doi.org/10.1016/0370-2693(88)90790-3}{Spergel et al.\ 88}
    &  {RTO} & \no           &  \yes&  \yes& \no           & \no           \\

\href{https://doi.org/10.1007/978-94-009-0921-2_8}{Chudakov et al.\ 88}
  & ---          & \no           & \yes          & \no           & \no           & \yes          \\

\href{https://doi.org/10.1103/PhysRevD.39.1013}{LoSecco 89}
  & ---          & \yes          & \yes          & \no           & \yes& \no           \\

\href{https://doi.org/10.1007/BF01549084}{Krivoruchenko 89}
  & ---          & \yes          & \yes          & \no           & \yes          & \no           \\

\href{https://doi.org/10.1111/j.1749-6632.1989.tb50547.x}{Loredo \& Lamb 89}
  & ITO          & \no           & \yes          & \yes          & \no           & \no           \\

\href{https://doi.org/10.1103/PhysRevD.41.2967}{Kie{\l}czewska 90}
  & ---          & \yes          & \yes          & \no           & \yes          & \no           \\

\href{https://doi.org/10.1103/PhysRevD.49.1389}{Smirnov et al.\ 93}
  & ---          & \no           & \yes          & \no           & \no           & \yes          \\

\href{https://doi.org/10.1016/0550-3213(94)00595-6}{Kernan et al.\ 95}
  & ITO          & \no           & \yes          & \yes          & \no           & \yes          \\

\href{https://doi.org/10.1103/PhysRevD.54.1194}{Jegerlehner et al.\ 96}
  & ---          & \no           & \yes          & \no           & \no           & \yes           \\

\href{https://doi.org/10.1103/PhysRevD.65.063002}{Loredo \& Lamb 02}
  & ITO          & \no           & \yes          & \yes          & \no           & \no           \\

\href{https://doi.org/10.1103/PhysRevD.70.043006}{Costantini et al.\ 04}
  & ---          & \yes          & \no          & \no           & \yes          & \yes          \\

\href{https://doi.org/10.1134/S1063773709010010}{Vissani et al.\ 08$^*$}
  & ---         & \yes          & \yes          & \no           & \yes          & \yes          \\

\href{https://doi.org/10.1016/j.astropartphys.2008.12.010}{Pagliaroli et al.\ 09$^*$}
  & ITO          & \no           & \yes          & \yes          & \yes          & \yes          \\

\href{https://doi.org/10.1103/PhysRevD.108.083040}{Fiorillo et al.\ 23}
  & ITO          & \no           & \yes          & \yes          & \no           & \no           \\

\href{https://doi.org/10.1140/epjc/s10052-023-11597-6}{Dedin Neto et al.\ 23$^*$}
  & ITO          & \no           & \yes          & \yes          & \yes          & \yes          \\

\href{https://doi.org/10.1016/j.physletb.2023.138252}{Volpe et al.\ 23}
  & ---          & \no           & \yes          & \no           & \no           & \yes          \\

\href{https://doi.org/10.1088/1475-7516/2025/05/027}{Bozza et al.\ 25}
  & ITO          & \no           & \yes          & \yes          & \yes          & \no           \\

\midrule
\textbf{This work 26$^*$} & {RTO} & \yes          & \yes          & \yes          & \yes          & \yes          \\
\bottomrule
\end{tabular}
\vspace{0.2cm}
\caption{\small Methodological matrix of SN1987A data analyses:
\yes~indicates inclusion, $\times$ indicates exclusion.
Asterisks indicate the studies that simultaneously integrated at least 4 physical items (described in the text).}
\label{tab:lit_comparison_final}
\end{table*}

Several key insights can be deduced from the chronological overview in Table \ref{tab:lit_comparison_final}:
\begin{itemize}
    \item \textbf{Gradual Progress:} The historical literature reveals a fragmented landscape, where researchers traditionally decoupled the temporal analysis from geometric or flavor-related parameters.
 \item \textbf{The RTO:} The difference in absolute time between IMB and Kamiokande-II---namely, a specific Relative Timing Offset---was initially introduced as an explicit clock parameter to be estimated in early analyses \cite{Burrows:1987zz,Abbott:1987bm,Spergel:1987ex} for the sole purpose of bounding neutrino masses. In all these papers
 1) The RTO parameter for Baksan relative to the IMB was not calculated;
 2) The error in the RTO parameters was never estimated.
 \item \textbf{The Shift to ITO:} Following \cite{Loredo:1988mk} and in all analyses of SN1987A conducted over the subsequent forty years, the RTO parameters have been abandoned and systematically replaced by the time interval between the arrival of the first neutrino and the detection of the event: the Intrinsic Timing Offsets.
  \item \textbf{Relevance of the Angular Distribution:} Examining the angular distribution is potentially critical in view of the anomalous forward-peaking historically observed in the data. Recall however that at the Yamada Meeting in 1988, Kamiokande-II revised the reconstructed direction of its second detected event, heavily biasing and softening the intense forward-directed data speculations previously surmised.
    \item \textbf{Non-Trivial Angular Integration:} The description of the observed angles of arrival within a rigorous $\chi^2$ statistical analysis remains notoriously challenging and technically demanding. This integration was attempted by only a few papers; among those, an even smaller fraction explicitly modeled the elastic scattering (ES) channel, and not all of them adopted conservative or theoretically sound assumptions for the underlying neutrino emission parameters.
\item \textbf{Studies of the Angular Distribution:} It should first be noted that some of the previous studies 1) account for the presence of ES events in the SN1987A dataset, but 2) do not analyze their angular distribution; and vice versa; only a subset does both.
To the best of our knowledge, the only previous attempt at a two-dimensional energy-angle analysis of the K1 event was presented in \cite{Vissani:2008wbj} (see Fig.~2 there).
\item \textbf{Methodological Isolation of ES:} Prior to the present study, works addressing the elastic scattering hypothesis either entirely neglected the temporal distribution or averaged it over the entire burst duration, stripping the signal of its dynamic imprint
 (e.g., this was done in \cite{Vissani:2008wbj}).
           \item \textbf{Modeling of Neutrino Flavor Evolution ($\nu$-osc):}
        After 1992 \cite{Pantaleone:1992eq,Sigl:1993ctk,Samuel:1993uw}
and most notably after \cite{Dolgov:2002ab,Hannestad:2006nj,Fogli:2007bk,Duan:2010bg},
    significant difficulties have emerged regarding the simulation of flavor conversion during the accretion and cooling phases, where non-linear neutrino-neutrino self-interactions can modify the MSW picture,
    triggering collective instabilities. This guided a profound rethinking of certain points that were thought to be well understood.
     \item
    \textbf{A Way Out from Uncertainties:}
    Crucially, our framework sidesteps these complications. Because our physical focus rests on the early neutronization phase, the standard, linear MSW matter-effect theory is fully sufficient and structurally robust to model the signal, providing an exceptionally clean statistical baseline.
        \item \textbf{Unprecedented Completeness:}
   Only a handful (three) of previous analyses achieved four simultaneous ``$\checkmark$'' in Table~\ref{tab:lit_comparison_final}. This work represents the unique formulation maximizing all five physical aspects.
\end{itemize}
Let us conclude by noticing that the prediction that MSW oscillations would dramatically alter a supernova signature---suppressing the sub-second prompt neutronization burst by the $\sigma_{\nu_\mu}/\sigma_{\nu_e}$ cross section ratio---was remarked by Walker \& Schramm (1987)
in a prescient study~\cite{Walker:1986xd},
appeared shortly after Mikheyev and Smirnov's landmark papers
\cite{Mikheyev:1985zog,Mikheev:1986wj}
and just a few
months before the historical observation of SN1987A. The present work honors that lineage by proceeding with the first high-precision, global $\chi^2$ analysis that simultaneously synchronizes the independent clocks via the RTO protocol and quantitatively bounds these pristine linear signatures.

\section{Conclusions}

We resolved the long-standing timing ambiguity of the SN1987A neutrino burst by determining quantitatively and accurately the Relative Timing Offsets (RTOs) of the Baksan and Kamiokande-II clocks with respect to the IMB absolute reference.
Our unified core-collapse chronology overcomes historical clock uncertainties, improving absolute timing precision by two orders of magnitude.
This alignment enables a rigorous characterization of the physical mechanisms driving individual events.
Rather than a simple recalibration,
this alignment demonstrates how a rigorous treatment of non-Gaussian correlations
enables a quantitative characterization of the physical mechanisms driving individual
events, providing a solid mathematical foundation where only qualitative expectations existed.

Specifically, we addressed the hypothesis that the initial Kamiokande-II event originated from electron elastic scattering during the NP---a scenario motivated by its observed angular distribution. While the literature has
traditionally accepted an inverse beta decay origin by consensus, our multi-dimensional
likelihood analysis provides the first formal, quantified bound by simultaneously
incorporating temporal, energy, angular response non-Gaussianities, and MSW
flavor oscillation frameworks.
However, our quantitative likelihood analysis favors an inverse beta decay origin in the accretion phase, with a likelihood ratio of 3--6, depending on the specific MSW oscillation scenario. In combination with the time-integrated analysis of the accretion+cooling signal in \cite{Vissani:2008wbj}, the computation of the temporal distribution in Sec.\ref{sec:es_hypothesis} constitutes the strongest support for the IBD interpretation of SN1987A data. We conclude that the angular anomaly cannot be explained by the presence of ES events.
Crucially, this result does not merely reiterate past consensus; rather, the primary strength of our framework lies in its unique ability to transform this long-standing, unquantified
assumption within the community into a mathematically rigorous and statistically bounded physical constraint.
The formal analysis proves that an ES origin during the prompt neutronization burst is subdominant under standard oscillation scenarios.

We employ parameterized model constrained by a maximum-likelihood analysis of the SN1987A data to conservatively account for uncertainties. Alternatively, this quantitative RTO-based procedure can be directly applied to simulated neutrino fluxes derived from numerical calculations, such as \cite{Fiorillo:2023frv}. Utilizing full numerical simulations may allow for an even more precise alignment of the Kamiokande-II and Baksan event timelines, depending on the residual astrophysical and oscillation uncertainties, particularly the signal rise time.

Ultimately, these results demonstrate that even with the sparse and unique dataset of SN1987A, sub-dominant signals can be rigorously and quantitatively constrained through a high-precision treatment of relative timing offsets.

For a future Galactic supernova, with significantly higher statistics, the likelihood analysis of the angular distribution presented here establishes an essential precision benchmark.
It will be crucial for the unambiguous identification of the neutronization burst and the achievement of millisecond timing precision.
Such a detection would offer an unprecedented window into the early stages of gravitational collapse and strengthen multi-messenger connections with gravitational-wave and electromagnetic signals.

\begin{acknowledgements}
The Authors thank G.~Matteucci and M.~Miceli for valuable discussions. V.d.R.~thanks the QTC at Danish IAS and IMADA (SDU) for hospitality.
G.R.~acknowledges support from INFN project ENP; F.V.~from PRIN 2022 grant 2022E2J4RK.
\end{acknowledgements}

\appendix

\section*{Appendixes}

\section{Calculation of the $\chi^2$-functions of the absolute times}
\label{AppendixA}
We detail the procedure to estimate the time $\bar{T}_0=T_i-t_i$ of the arrival of the first antineutrino (not detected) at the IMB location 
The $\chi^2$-function of $t_i$ is reported in figure 5 of \cite{Bozza:2025wqo} and in tabular form in \cite{SN1987Acodes}. The associated probability distribution function is given by $ \mathcal{L}(t_i)=\text{exp}\big[-\chi^2(t_i)/2\big]\,$.
The measured IMB time $T^{\text{exp}}_i$ is $7:35:41.374\pm 0.050\,[\textbf{UT}]$ and the probability distribution function is a Gaussian $G(T_i)$.
Since $T_{i}$ and $t_i$ are independent, the probability distribution function for $\bar{T}_0$ is simply given by
\begin{equation}
    \mathcal{L}(\bar{T}_0)=\int dt_i\,G(\bar{T}_0+t_i)\times \mathcal{L}(t_i)\,.
\end{equation}
The function $ \mathcal{L}(\bar{T}_0)$ has been estimated numerically and the corresponding $\chi^2$-function is $\chi^2(\bar{T}_0)=-2\log{\mathcal{L}(\bar{T}_0)}$, from which we recover the best fit value and the  $1\sigma$ confidence interval. 

An analogous procedure can be applied to estimate the absolute times of detection of the first antineutrinos in Kamiokande-II and Baksan, $\bar{T}_{k(b)}$, exploiting the $\chi^2-$functions of the RTOs in Fig.\,\ref{fig:time_diff_delay}. The associated probability distribution function is $\mathcal{L}(\Delta t_{k(b)})=\text{exp}\big[-\chi^2(\Delta t_{k(b)})/2\big]$.
Neglecting Earth's size, $\bar{T}_{k(b)}=T_i+\Delta t_{k(b)}$ and its probability distribution function is 
\begin{equation}
    \mathcal{L}(\bar{T}_{k(b)})=\int d\,\Delta t_{k(b)}\, G(\bar{T}_{k(b)}-\Delta t_{k(b)})\times \mathcal{L}(\Delta t_{k(b)}).
\end{equation}
 The $\chi^2$-function of $\bar{T}_{k(b)}$ is defined as $\chi^2(\bar{T}_{k(b)})=-2\log \mathcal{L}(\bar{T}_{k(b)})$, from which we recover the best-fit value and the $1\sigma$ confidence interval. 
 The $\chi^2$ tables corresponding to Fig.~\ref{fig:time_diff_delay} and Fig.~\ref{fig:times_comparison}  
and the  codes to calculate Fig.~\ref{fig:K1comparison} are given in reference~\cite{SN1987Acodes}. The variation of the best fit of RTOs varying $t_{\text{max}}$ is of order $10^{-2}\,\mbox{s}$ as shown in Table~\ref{tab:value_tt}.

\begin{figure*}[t]
\centering
\includegraphics[width=0.5\textwidth]{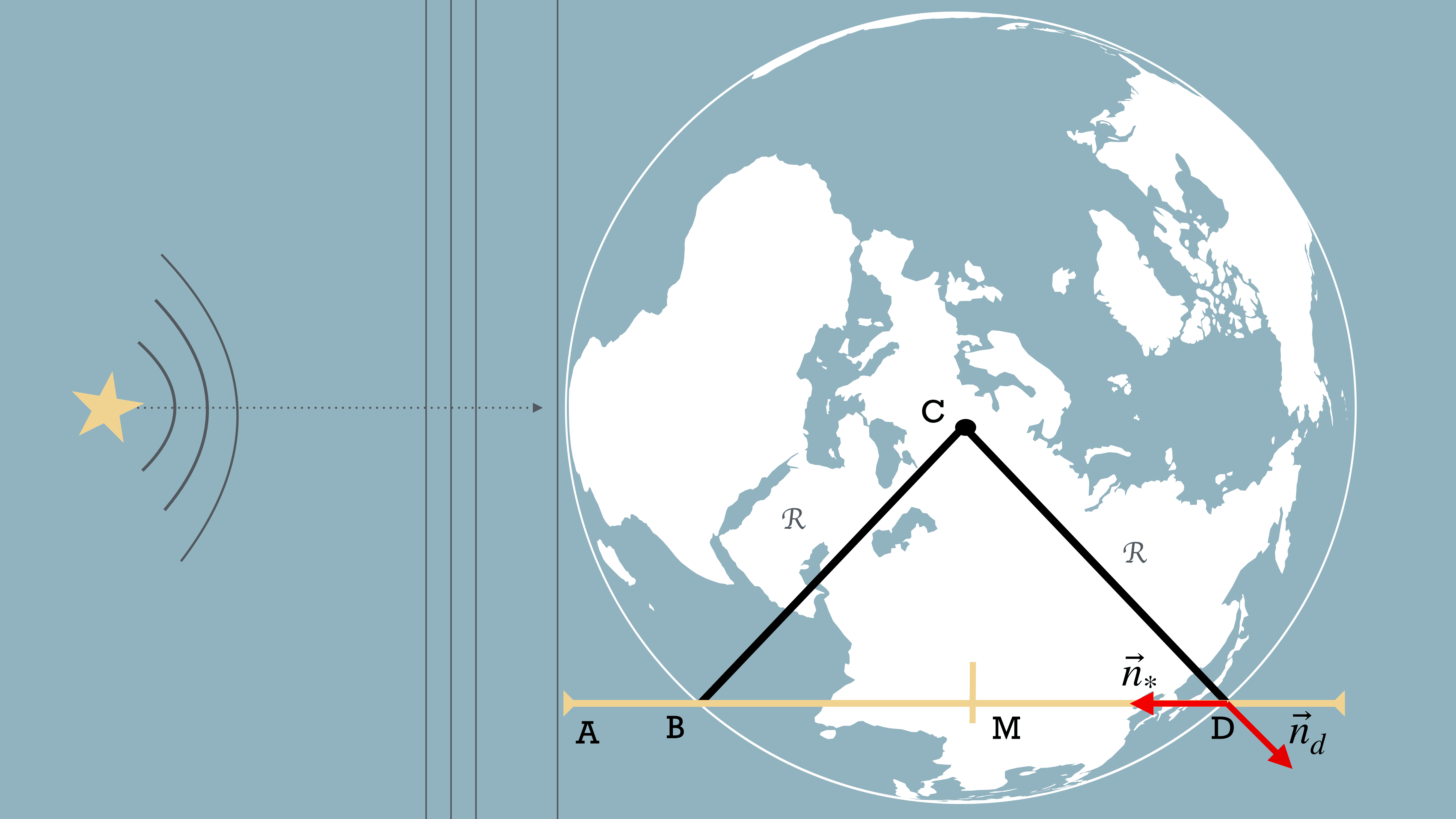}
        \caption{\small Relative positions of SN1987A 
        and detectors on Earth. The unit vectors $\vec{n}_d$ and $\vec{n}_*$ are also shown.}
        \label{fig:earth}
\end{figure*}

\section{Detector positions and geometry}
\label{AppendixB}
The neutrinos propagated through the Earth's interior before reaching the detectors, as the SN1987A progenitor was located in the Southern Sky and all three detectors were situated in the Northern Hemisphere.

We adopt the IMB time and location as a global reference. 
The distance traveled within the Earth for each site $d$ is given by $L_d = -2\mathcal{R} \langle\vec{n}_d, \vec{n}_*\rangle$, where $\mathcal{R} = 6371$~km is the Earth's mean radius,
see Fig.~\ref{fig:earth}.
In geocentric celestial coordinates, any unit vector is defined as $\vec{n} = (\cos\delta \cos\alpha, \cos\delta \sin\alpha, \sin\delta)$, where $\delta$ is the declination and $\alpha$ is the Right Ascension (RA)   (not be confused with the ``pinching" parameter in Eq.~\ref{pinch}).
For the detector unit vectors $\vec{n}_d$, the declination corresponds to the geographic latitude $\lambda$, while the RA is derived from the Local Sidereal Time (LST) at the moment of detection $T_i$. The coordinates for the SN1987A progenitor and the respective detectors used in this work are summarized in Table~\ref{Table:position}, along with the calculated chord lengths $L_d$.




\bibliographystyle{spphys} 


\bibliography{bib.bib}

\end{document}